\documentclass[preprint]{aastex}
\singlespace
\usepackage{latexsym}
\usepackage{fullpage}
\usepackage{graphics}
\usepackage{graphicx}
\usepackage{rotating}
\usepackage{epsfig}
\usepackage{psfig}

\begin{document}
\title{Swift Pointing and the Association Between 
Gamma-Ray Bursts and Gravitational-Wave Bursts}


\author{Lee Samuel Finn,\thanks{Also Center for
Gravitational Physics and Geometry, Department
of Physics, and Department of Astronomy and Astrophysics;
\texttt{e-mail: lsf@gravity.psu.edu}}~ 
Patrick J. Sutton\thanks{Also Center for Gravitational Physics and
Geometry and Department of Physics; \texttt{e-mail: psutton@gravity.psu.edu}}
}
\affil{Center for Gravitational Wave Phenomenology, Department of
Physics,\\The Pennsylvania State University, University Park, PA
16802, USA }
\and
\author{Badri Krishnan\thanks{Also Center for Gravitational Wave
Phenomenology and Center for Gravitational Physics and Geometry,
Pennsylvania State University, University Park, PA-16801, USA; 
\texttt{e-mail: badkri@aei.mpg.de}}}
\affil{Max Planck Institut f\"ur Gravitationsphysik,\\ Am M\"uhlenberg
1, 14476 Golm, Germany}

\begin{abstract}
  The currently accepted model for gamma-ray burst phenomena involves
  the violent formation of a rapidly rotating solar mass
  black hole. Gravitational waves should be associated with the black-hole 
  formation, and their detection would permit this model to be
  tested, the black hole progenitor (e.g., coalescing binary or collapsing 
  stellar core) identified, and the origin of the
  gamma rays (within the expanding relativistic fireball or at the
  point of impact on the interstellar medium) located. 
  Even upper limits on the gravitational-wave strength associated 
  with gamma-ray bursts could constrain the gamma-ray burst model.
  To do any of these requires joint observations of gamma-ray
  burst events with gravitational and gamma-ray detectors. Here we
  examine how the quality of an upper limit on the gravitational-wave
  strength associated with gamma-ray burst observations depends on the
  relative orientation of the gamma-ray-burst and gravitational-wave 
  detectors, and apply our results to the particular case of the
  Swift Burst-Alert Telescope (BAT) and the LIGO gravitational-wave
  detectors. A result of this investigation is a science-based
  ``figure of merit'' that can be used, together with other
  mission constraints, to optimize the pointing of the Swift telescope
  for the detection of gravitational waves associated with gamma-ray
  bursts. 


\end{abstract}

\section{Introduction}
\label{sec:Introduction}

The currently accepted model for gamma-ray burst phenomena
involves the violent formation of an approximately solar-mass black
hole surrounded by a similarly massive debris torus.  The gamma-ray
burst is powered by the release of binding energy as the debris torus
is accreted onto the black hole.  The formation of the black hole and
debris torus may take place through the coalescence of a compact
binary or the collapse of a quickly rotating massive stellar core; 
the gamma-ray burst emission may 
take place at the site of crossing internal shock waves within the
expanding relativistic fireball powered by the accretion onto the
black hole, or as the fireball is decelerated by the interstellar
medium \citep{MeRe:93, ReMe:94}. 

A gravitational-wave burst is likely to be associated with the
formation of the rapidly rotating, central black-hole engine of a 
gamma-ray burst. 
The character of the gravitational-wave burst (e.g.,~energy,
spectrum, polarization) will depend on the degree of non-axisymmetry
associated with the collapse, and thus its progenitor 
\citep{KoMe:02,KoMe:03}.  Similarly, the relative time of arrival 
of the gravitational and gamma-ray bursts
will depend on whether the gamma-ray burst is generated by
internal shocks in the exploding fireball or external shocks when the
fireball is decelerated by the interstellar medium.\footnote{
Though the fireball is highly relativistic, it travels subluminally; 
hence, the further the gamma rays are produced from the central engine 
the longer the time lag between the arrival of the gravitational waves 
and the gamma rays.  This time lag can range from 1 second to 100 seconds.
On the other hand, the time lag between the black-hole formation (which produces the 
gravitational-wave burst) and the launch of the relativistic jet 
(which produces the GRB) should be on the order of the dynamical 
timescale of milliseconds, and is a much smaller effect.}  
Observation of
gravitational-wave bursts associated with gamma-ray bursts thus may
reveal details of the gamma-ray burst mechanism that cannot be
revealed through observations of the gamma rays alone. 
In this paper we examine how gravitational-wave observations using 
the LIGO detectors \citep{Si:01} can be made in conjunction with 
gamma-ray burst observations
by the Swift satellite to determine whether there
is an association between gamma-ray and gravitational-wave bursts, a
first step in the ambitious program of developing gravitational-wave
observations into a tool of astronomical discovery.

\citet{FiMoRo:99} have described how the cross-correlated output of
two gravitational-wave detectors, taken in coincidence with
gamma-ray burst (GRB) events, can be used to detect or place upper
limits on the emission of gravitational-wave bursts (GWBs) by GRBs.
This procedure has already been used in the analysis of data
from the EXPLORER and NAUTILUS gravitational-wave detectors at times
associated with 47 GRBs detected by the BeppoSAX satellite to bound,
at 95\% confidence, the root-mean-square gravitational-wave strain 
$h_{\mbox{\tiny{}RMS}}$ associated with these GRBs to less than
$6.5\times10^{-19}$ in the gravitational-wave-detector waveband
(approximately 0.5 Hz about 900 Hz), assuming that the gamma rays
originate in internal shocks \citep{As:02}.  
\citet{FiMoRo:99} estimate that 1000 GRB observations
combined with observations from the (broad-band) initial LIGO detectors could
produce an upper limit on the gravitational-wave strain associated
with GRBs of approximately $h_{\mbox{\tiny{}RMS}} \le
1.7\times10^{-22}$ at 95\% confidence.

An important consequence of the cosmological origin of GRBs is their
isotropic distribution on the sky.  In their original work 
\citet{FiMoRo:99} assumed that GRBs would be detected isotropically 
as well -- i.e.,~that the GRB detector had an isotropic antenna pattern.
They did note, however,
that the Swift satellite, a next-generation multi-wavelength satellite
dedicated to the study of GRBs, does not have an isotropic antenna
pattern and that this has potentially important consequences for the
ability of the combined gamma-ray burst/gravitational-wave detector
array to detect or limit the gravitational-wave flux on Earth owing to
GRBs.  Here we study this question specifically in the context of the
Swift satellite and the LIGO gravitational-wave detectors; i.e.,~we
determine, as a function of Swift's pointing, the sensitivity
of the Swift/LIGO detector array to gravitational waves from GRBs, and
propose a figure-of-merit that can be used in Swift mission scheduling
to optimize the sensitivity of the Swift/LIGO array to the
gravitational-wave flux from GRBs.  We find that the upper limit that
can be placed on $h_{\mbox{\tiny{}RMS}}$ differs by a factor of 2
between best and worst orientations of the satellite.

We begin in Section~2 with a review of how \citet{FiMoRo:99} proposed
using Student's $t$-test to detect a GRB--GWB association and place an
upper limit on the gravitational-wave strength associated with GRBs.
We extract the direction dependence of this upper limit in Section~3,
and apply to the case of Swift in Section~4.  We conclude with some
brief remarks in Section~5.

\section{Observing a GRB--GWB association}

\citet{FiMoRo:99} described how the signal from 
two independent gravitational-wave detectors can be analyzed to 
identify the gravitational-wave signal associated with gamma-ray 
bursts and either bound or measure the population average of the 
gravitational-wave flux on Earth from this potential source.
In this section we review their analysis methodology in anticipation 
of using it to determine the sensitivity of joint LIGO/Swift observations 
to the detectors' relative orientation.

\subsection{Detecting a GRB--GWB association}

Consider a set of $N$ GRB detections.  Assume that, as a result of
each detection, we know the direction to the source
$\widehat{\Omega}_k$ and the arrival time ${\tau}_k$ of the burst at
Earth's barycenter.  For our purposes, each GRB observation is
completely characterized by the pair $(\widehat{\Omega}_k,\tau_k)$.\footnote{
Afterglow observations will also give the distance to the GRB source  
\citep{Me_etal:97}.}  
Focus attention on a pair of gravitational-wave detectors located
at positions $\mathcal{D}_i$ ($i = 1, 2$) relative to Earth's
barycenter. The arrival time of GRB $k$ at detector $i$ is
\begin{equation}\label{eq:tn}
t^{(i)}_k = \tau_k - \widehat{\Omega}_k\cdot\mathcal{D}_i
\end{equation}
in units where the speed of light $c$ is unity. 

\citet{FiMoRo:99} note that the two LIGO detectors are very nearly
co-planar and co-aligned.  Consequently, a plane gravitational wave
incident on the detector pair from the direction $\widehat{\Omega}_k$
will lead to correlated detector responses with a time lag equal to 
\begin{equation}
\Delta t_k = t^{(2)}_k - t^{(1)}_k.
\end{equation}
To identify the presence of GWBs 
associated with GRBs, \citet{FiMoRo:99} focus attention on the
correlated energy in the detector outputs corresponding to plane GWBs
incident on the detectors from the direction of the corresponding
GRBs: i.e., the correlation of the detector outputs at times that
differ by $\Delta t_k$.

Let $s_i(t)$ be the output of gravitational wave detector
$\mathcal{D}_i$, which we assume to consist of detector noise $n_i(t)$
and a possible gravitational wave signal $h_i(t)$ {\em produced by the
GRB source}:\footnote{There may be a flux of gravitational waves from
  other sources incident coincidentally on the detector at the same
  time; however, since this radiation is not correlated with the
  gamma-ray burst it is, for our purpose, noise and we lump its
  contribution to $s$ in with $n$.}
\begin{equation}\label{eq:s} 
s_i(t) = n_i(t) + h_i(t) \, .
\end{equation}
\citet{FiMoRo:99} define 
\begin{equation}
\label{eq:cross} 
S(\widehat{\Omega}_k,\tau_k) 
=  \int_0^T \!\!dt \int_0^T \!\!dt' \, s_{1}(t_k^{(1)}-t)
          \,Q(t-t')\, s_{2}(t_k^{(2)}-t') \, ,
\end{equation}
as the energy in the cross-correlation of the
two detectors corresponding to the GRB characterized by
$(\widehat{\Omega}_k,\tau_k)$. 
Here $Q$ is a freely specifiable, symmetric filter function
\citep{FiMoRo:99}, and $T$ is chosen large enough to encompass the
range of possible times by which the gravitational waves from a GRB
event may precede the gamma rays, which is typically thought to be of
order 1 s for GRBs produced by internal shocks and 100 s for GRBs
produced by external shocks \citep{SaPi:97, KoPiSa:97}. 

Writing the detector output $s_i$ as the sum of
the detector noise $n_i$ and the gravitational-wave signal $h_i$ 
associated with the GRB we can, in turn, write 
\begin{equation}\label{eq:S}
S_k \equiv S(\widehat{\Omega}_k,\tau_k)  
  =  \langle n_1,n_2 \rangle +  \langle n_1,h_2 \rangle + 
     \langle h_1,n_2 \rangle +  \langle h_1,h_2 \rangle 
     \, , 
\end{equation}
where 
\begin{equation}
\langle f,g \rangle 
  =  \int_0^T \!\!dt \int_0^T \!\!dt' \, f(t_k^1-t)
          \,Q(t-t')\, g(t_k^2-t') \, .
\end{equation}
The terms $\left<n_i,h_j\right>$ in equation (\ref{eq:S}) vanish in 
the mean over an ensemble of noise since
the noise in our gravitational wave detector is uncorrelated with any
gravitational wave signal. The term $\left<n_1,n_2\right>$ is, in the 
noise ensemble mean, a
constant, which will be zero if the noise in the two detectors is
uncorrelated. 

All four of the contributions to $S$ in equation (\ref{eq:S}) are
generally unknown for any particular GRB. 
Correspondingly, \citet{FiMoRo:99} consider the collection
$\widehat{\mathcal{S}}_{\mbox{\tiny on}}$ of $S_k$, 
\begin{equation}
\widehat{\mathcal{S}}_{\mbox{\tiny on}} = \left\{S_k: k = 1\ldots N\right\} \, ,
\end{equation}
and a second collection $\widehat{\mathcal{S}}_{\mbox{\tiny off}}$
\begin{equation}
\widehat{\mathcal{S}}_{\mbox{\tiny off}} = \left\{S'_m: m = 1\ldots M\right\}
\, , 
\end{equation}
where each $S'_m$ is constructed as in equation (\ref{eq:cross}) but with
a $(\widehat{\Omega},\tau)$ pair chosen randomly and not associated
with a GRB. The collections $\widehat{\mathcal{S}}_{\mbox{\tiny on}}$
and $\widehat{\mathcal{S}}_{\mbox{\tiny off}}$ are samples drawn from
{\em populations} ${\mathcal{S}}_{\mbox{\tiny on}}$ and
${\mathcal{S}}_{\mbox{\tiny off}}$. The sample means
$\widehat{\mu}_{\mbox{\tiny on}}$ and $\widehat{\mu}_{\mbox{\tiny
    off}}$ and variances $\widehat{\sigma}^2_{\mbox{\tiny on}}$ and
$\widehat{\sigma}^2_{\mbox{\tiny off}}$ are estimates of the
{\em population means} ${\mu}_{\mbox{\tiny on}}$ and ${\mu}_{\mbox{\tiny
    off}}$ and variances ${\sigma}^2_{\mbox{\tiny on}}$ and
${\sigma}^2_{\mbox{\tiny off}}$. These are, in turn, related by 
\begin{eqnarray}
\mu_{\mbox{\tiny on}} - \mu_{\mbox{\tiny off}} &=& 
\overline{\left<h_1,h_2\right>} \, , \\
\sigma^2_{\mbox{\tiny on}} - \sigma^2_{\mbox{\tiny off}} &=& 
\overline{\left<n_1,h_2\right>^2} + 
\overline{\left<n_2,h_1\right>^2} + \mathcal{O}(h^3) \, ,
\end{eqnarray}
where the overbar represents a mean over the population of GRBs. 
Comparing the two sample sets $\widehat{\mathcal{S}}_{\mbox{\tiny
    on}}$ and $\widehat{\mathcal{S}}_{\mbox{\tiny off}}$ thus provides
a sensitive measure of the presence or absence of gravitational waves
associated with GRBs. 

When the detector noise is sufficiently well-behaved that terms like
$\left<n_1,h_2\right>$, $\left<n_2,h_1\right>$, and 
$\left<n_1,n_2\right>$ are normally distributed the Student $t$-test 
\citep{SnCo:67} can be used to distinguish between the two sample
sets; in other cases a non-parametric test such as the Mann-Whitney
test \citep{SiCa:88} can be used.

In the Student $t$-test the difference between the two distributions
represented by the sample sets $\widehat{S}_{\mbox{\tiny{}on}}$ and
$\widehat{S}_{\mbox{\tiny{}off}}$ is characterized by the
$t$-statistic:
\begin{eqnarray}
\widehat{t}  & = &  
\frac{\widehat{\mu}_{\mbox{\tiny{}on}}-\widehat{\mu}_{\mbox{\tiny{}off}}}{\widehat{\Sigma}}
\sqrt{\frac{N_{\mbox{\tiny{}on}}N_{\mbox{\tiny{}off}}}{N_{\mbox{\tiny{}on}}
    +N_{\mbox{\tiny{}off}}}}
\, , \label{eq:tstat} \\ 
\widehat{\Sigma}^2  & = &  
\frac{\widehat{\sigma}_{\mbox{\tiny{}on}}^2(N_{\mbox{\tiny{}on}}-1)
  +\widehat{\sigma}_{\mbox{\tiny{}off}}^2(N_{\mbox{\tiny{}off}}-1)}{N_{\mbox{\tiny{}on}}+ 
  N_{\mbox{\tiny{}off}}-2}
\, , \label{eq:Sigma} 
\end{eqnarray}
If both $\widehat{\mathcal{S}}_{\mbox{\tiny{}on}}$ and
$\widehat{\mathcal{S}}_{\mbox{\tiny{}off}}$ are drawn from same
normal distribution then the distribution of $\widehat{t}$ is given by
Student's distribution with
$N_{\mbox{\tiny{}on}}+N_{\mbox{\tiny{}off}}-2$ degrees of freedom 
\citep{Cr:99}. This distribution itself tends toward a normal
distribution with unit variance when
$N_{\mbox{\tiny{}on}}+N_{\mbox{\tiny{}off}}$ is large.

Now suppose that there is no GWB--GRB association.  In this event
$\widehat{S}_{\mbox{\tiny{}on}}$ and $\widehat{S}_{\mbox{\tiny{}off}}$
are drawn from the same distribution and there is a number $t_0(p)$
such that $\widehat{t}$ will be less than $t_0(p)$ in a fraction $p$
of all observations of sample sets $\widehat{S}_{\mbox{\tiny{}on}}$
and $\widehat{S}_{\mbox{\tiny{}off}}$.\footnote{ On physical grounds
  the expectation value of $\widehat{t}$ will be positive
  semi-definite for the LIGO detector pair if gravitational waves are
  associated with gamma-ray bursts.}  If, in our particular
observation of $\widehat{S}_{\mbox{\tiny{}on}}$ and
$\widehat{S}_{\mbox{\tiny{}off}}$ $\widehat{t}$ is less that $t_0(p)$
then we accept the hypothesis that there are {\em no\/}
gravitational-waves associated with gamma-ray bursts.  If, on the
other hand, we find $\widehat{t}$ greater than $t_0(p)$ then we
reject, with confidence $p$, this hypothesis; i.e., we assert that
there is an association of gravitational-waves with gamma-ray bursts.

\subsection{Setting an upper limit on the gravitational-wave strength 
associated with GRBs}

As described in the previous section, Student's $t$-test tells us only if there
is a link between GWBs and GRBs.  An alternative analysis, also
described by \citet{FiMoRo:99}, uses the $t$ statistic to derive
a confidence interval or upper limit on the population-averaged
gravitational-wave flux associated with GRBs from a measured value
$\widehat{t}$ of the $t$ statistic.  In this analysis we derive the
classical confidence interval or upper limit from the probability
distribution $P(t|h,I)$ of the $t$ statistic assuming that GRBs
radiate GWBs with intrinsic amplitude described by $h$ and other model
parameters (gravitational-wave burst luminosity function, burst
characteristic, etc.) described by $I$.

In the limit that the GWBs are weak relative to the sensitivity of the
individual gravitational-wave detectors and the numbers of on- and
off-source observations are separately large, 
\citet{FiMoRo:99} showed that
\begin{equation}
P(t|h,I) = N\left(t\left|\mu,\frac{\sigma^2}{{2}}\right.\right) \, ,
\end{equation}
where 
\begin{eqnarray}
\mu  
  & = &  \overline{\langle h_1,h_2 \rangle}  \\
\sigma^2  
  & = &  \frac{T}{4}\int_{-\infty}^{\infty}\!\!df 
         P_{1}(f)P_2(f) |\widetilde{Q}(f)|^2 \, .
\end{eqnarray}
Here $P_i(f)$ is the one-sided power spectral density of the
$i^{\mathrm{th}}$ detector, defined as 
\begin{equation}
P_i(|f|) = 2 \int_{-\infty}^{\infty} \!dt\, e^{i2\pi ft}
           n_i(\tau) n_i(\tau+t) \, ,
\end{equation}
$\overline{\langle h_1,h_2 \rangle}$ is the average of $\langle
h_1,h_2 \rangle$ over the GRB population and the associated GWB
luminosity function described by $I$, $N(t|\mu,\nu)$ is the normal
distribution with mean $\mu$ and variance $\nu$, and
$\widetilde{Q}(f)$ is the Fourier transform of $Q(\tau)$.  For
larger-amplitude GRBs or different sample sizes (e.g.,~smaller number
of GRB observations) the distribution can be determined via Monte
Carlo simulations.  An observation of $t$ thus allows us to find a
confidence limit on $\overline{\langle h_1,h_2 \rangle}$ \citep{FeCo:98}, which
describes the flux of gravitational-waves on Earth owing to GRBs.

\section{Direction-dependence of the upper limit}

The analysis described by \citet{FiMoRo:99} assumed that the arms of
the two LIGO gravitational wave detectors all resided in the same
plane, that pairs of arms were parallel to each other, and that the
antenna pattern of the detectors was isotropic on the sky. In this
section we relax all of these approximations: i.e., we properly
account for the position and orientation of the two LIGO detectors on
the Earth and the dependence of their sensitivity to the direction to
the GRB source. Our result is an expression for the dependence of the
upper limit on population-averaged gravitational-wave strength
$\overline{\left<h_1,h_2\right>}$ as a function of the distribution 
of \emph{detected} GRBs on the sky. 
In section \ref{sec:fom} we combine this result with
the directional sensitivity of the Swift detector to determine the
dependence on Swift pointing of the upper limit on
$\overline{\left<h_1,h_2\right>}$ that can be set by joint LIGO/Swift
observations.

The gravitational-wave component $h_i$ of the LIGO detector output is,
in the small antenna limit,\footnote{This is appropriate for
  gravitational wave frequencies in the LIGO detector band.} a linear
function of the physical gravitational-wave strain
$h_{ab}(t,\vec{x})$,
\begin{equation}\label{eq:h}
h_i(t) = h_{ab}(t,\vec{x}_i)\, d^{ab}_i \, ,
\end{equation}
where $\vec{x}_i$ is the gravitational-wave detector's location. For
interferometer $i$ with arms pointing in the
directions $\widehat{X}_i$, $\widehat{Y}_i$, 
\begin{equation}\label{eq:dIFO}
d^{ab}_i = \frac{1}{2} ( \widehat{X}^a_i \widehat{X}^b_i
  - \widehat{Y}^a_i \widehat{Y}^b_i ) \, .
\end{equation}
(With this normalization $h_i(t)$ is equal to the fractional change in
differential arm length.) For gravitational wave bursts incident on
the Earth from direction $\widehat{\Omega}$,  
\begin{equation}\label{eq:tshift}
h_{ab}(t_k^{(1)}-t,\vec{x}_1) =
h_{ab}(t_k^{(2)}-t,\vec{x}_2) = h_{ab}(\tau_k-t,\vec{0})
\end{equation}
where $t_k^{(i)}$ is defined in terms of the direction to the source
$\widehat{\Omega}$ by equation (\ref{eq:tn}).  We can thus ignore the
physical separation of the detectors when computing the
cross-correlation statistic (cf.\ equation (\ref{eq:cross})). Finally, it is
convenient to resolve $h_{ab}$ on the two polarization tensors
$\epsilon^{+}$ and $\epsilon^{\times}$,
\begin{equation}\label{eq:hdecomp}
h_{ab}(t,\widehat{\Omega})  =  
h_+(t)\epsilon^+_{ab}(\widehat{\Omega}) 
+ h_\times(t) \epsilon^\times_{ab}(\widehat{\Omega}) \, ,
\end{equation}
where 
\begin{equation}
\epsilon^{+}:\epsilon^{+}      ~=~ 
\epsilon^{\times}:\epsilon^{\times} ~=~  2 \, ,
\end{equation}
\begin{equation}
\epsilon^{+}:\epsilon^{\times}  =  0 \, ,
\end{equation}
and
\begin{equation}
\epsilon^{+}\cdot\widehat{\Omega} =
\epsilon^{\times}\cdot\widehat{\Omega} = 0 \, .
\end{equation}
Lacking any detailed model for the gravitational waves that may
be produced in a GRB event, we make the following assumptions about
$h_{+}$ and $h_{\times}$:
\begin{itemize}
\item The waves have equal power in the two polarizations:
\begin{equation}\label{eq:power} 
\overline{h_+(t) h_+(t')} = \overline{h_\times(t) h_\times(t')} \, .
\end{equation}

\item The two polarizations are uncorrelated:
\begin{equation}\label{eq:uncorrel}  
\overline{h_+(t) h_\times(t')} = 0 \, . 
\end{equation} 
\end{itemize}

Focus attention now on the mean gravitational-wave contribution
$\overline{\left<h_1,h_2\right>}$ to the $t$ statistic. This
contribution depends on both the gravitational wave detector
sensitivity to GWBs arriving from different directions
as well as the gamma-ray bursts detector sensitivity to GRBs from
different directions (which determines the relative number of bursts
that will be observed from that direction). For GWBs
arriving from the direction $\widehat{\Omega}$, 
\begin{equation}
\left<h_1,h_2\right> 
  =  \rho_{\mbox{\tiny{}GWB}}(\widehat{\Omega}|d_1,d_2)
     \int_0^T\!\!dt\! \int_0^T\!\!dt'\,Q(t-t')\,h_{+}(t)\,h_{+}(t')\,,
\end{equation}
where 
\begin{equation}\label{eq:rhoLIGO} 
\rho_{\mbox{\tiny{}GWB}}(\widehat{\Omega}|d_1,d_2)
  \equiv  \sum_{A=+,\times}  
          \left(d_1 : \epsilon^A(\widehat{\Omega})\right)\,
          \left(d_2 : \epsilon^A(\widehat{\Omega})\right) 
\end{equation}
describes the direction-dependence of the sensitivity of the 
gravitational-wave detector pair to the GWB (cf.\ equations (\ref{eq:cross}),
(\ref{eq:power}), and (\ref{eq:uncorrel})).

To complete the evaluation of $\overline{\left<h_1,h_2\right>}$ turn
to the fraction of GRB detections that arise from different
patches on the sky. Since the intrinsic GRB population is isotropic,
the distribution of detection on the sky depends entirely on the
directional sensitivity of the GRB detector. Let the fraction of 
GRB detections in a sky patch of area $d^2\Omega$ centered at
$\widehat{\Omega}$ be given by 
\begin{equation}
\rho_{\mbox{\tiny{}GRB}}(\widehat{\Omega}|\widehat{\Omega}',\widehat{n})
d^2\widehat{\Omega}\,,
\end{equation}
where the GRB detector orientation is given by $\widehat{\Omega}'$,
the direction in which the detector is pointed, and $\widehat{n}$, 
which describes the rotation of the satellite about its pointing direction.

In terms of $\rho_{\mbox{\tiny{}GWB}}$ and
$\rho_{\mbox{\tiny{}GRB}}$ the mean gravitational wave contribution
$\overline{\left<h_1,h_2\right>}$ to $t$ for a gamma-ray burst
detector with fixed orientation $(\widehat{\Omega}',\widehat{n})$ is thus
\begin{equation}
\overline{\left<h_1,h_2\right>} =
\left[\int d^2\Omega\,\,
\rho_{\mbox{\tiny{}GWB}}(\widehat{\Omega}|d_1,d_2)
\rho_{\mbox{\tiny{}GRB}}(\widehat{\Omega}|\widehat{\Omega}',\widehat{n'})
\right]\left[
\int_0^T\!dt'\!\int_0^T\!dt\,Q(t-t')h_{+}(t)h_{+}(t')
\right]
\end{equation}
The first bracketed term contains all the direction and orientation
dependence of the gravitational wave and gamma-ray burst detectors,
while the second term is strictly a property of the gravitational
waves without reference to the orientation of the
detectors. Correspondingly, the sensitivity of the upper limit on
gravitational wave strength averaged over the observed GRB population 
when the orientation of GWB and the GRB detectors are
given by $(d_1, d_2, \widehat{\Omega}, \widehat{n})$ is
proportional to 
\begin{equation}\label{eq:FOM}
  \zeta(\widehat{\Omega}, \widehat{n},d_1, d_2)
  =  
  \int \!d^2\widehat{\Omega}'\,
  \rho_{\mbox{\tiny{}GRB}}(\widehat{\Omega}'|\widehat{\Omega},\widehat{n})\,
  \rho_{\mbox{\tiny{}GWB}}(\widehat{\Omega}'|d_1, d_2) 
  \, .
\end{equation}

Satellite GRB detectors orbit the Earth and so their orientation is
constantly changing; similarly, the orientation of Earth-based
gravitational wave detectors are constantly changing as the Earth
rotates about its axis. The quantity $\overline{\left<h_1,h_2\right>}$ will, in
the end, involve the time average of $\zeta$ over all these motions.
Since our principal purpose here is to evaluate the sensitivity of the
GRB/GWB detector array to GWBs from GRBs as a function of the relative
orientations of the detectors we focus on $\zeta$.

Clearly $\zeta$ can be regarded as a figure of merit that describes
how capable the gravitational-wave/gamma-ray burst detector
combination is at identifying GWBs associated with GRBs as a function
of the detector orientations. This figure of merit may be normalized
to have a maximum of unity; however, regardless of the normalization
\begin{equation}
  \zeta(\widehat{\Omega}',\widehat{n}', d_1, d_2) /
  \zeta(\widehat{\Omega},\widehat{n}, d_1, d_2) 
\end{equation}
is the ratio of the upper limits on the squared gravitational-wave
amplitude that can be attained by orienting the GRB satellite as
$(\widehat{\Omega},\widehat{n})$ versus
$(\widehat{\Omega}',\widehat{n}')$. To the extent that, e.g., the GRB
detector orientation $(\widehat{\Omega},\widehat{n})$ can be
manipulated on orbit, choosing orientations that maximize $\zeta$ will
lead to larger signal contributions to $t$ and thus more sensitive measurements of the
gravitational wave strength associated with GRBs.

\section{LIGO and Swift}\label{sec:fom}

Let us now consider the special case of the Burst-Alert Telescope
(BAT) on the Swift satellite\footnote{
http://swift.gsfc.nasa.gov/
}  
and the LIGO gravitational
wave detectors \citep{Si:01}.  The BAT is a wide field-of-view
coded-aperture gamma-ray imager that will detect and locate GRBs with
arc-minute positional accuracy. Its sensitivity to GRBs depends on the
the angle $\lambda$ between the line of sight to the GRB and the BAT
axis, as well as the rotational orientation of the satellite about the BAT axis.  
Averaged over this azimuthal angle, the BAT sensitivity as a function
of $\lambda$ is approximately\footnote{
See the BAT section of the Swift homepage,  
http://swift.gsfc.nasa.gov/science/instruments/bat.html  
} 
\begin{equation}\label{eq:rhoSwift}
\rho_{\mbox{\tiny{}GRB}}
  =  \left\{
       \begin{array}{ll}
         2\cos\lambda -1 + 0.077\sin{[13(1-\cos\lambda)]} & 
         \lambda \in [0,\pi/3] \, , \\
         0 & \mbox{otherwise} \, . 
       \end{array}
     \right.
\end{equation}
For the purposes of illustration we will
use this azimuthal-angle averaged expression for the BAT sensitivity. 

The second component of $\zeta$ is the function
$\rho_{\mbox{\tiny{GWB}}}$, which depends on the projection of the
gravitational-wave strain $h_{ab}$ on to the gravitational-wave
detector (cf.\ equations (\ref{eq:h}) and (\ref{eq:hdecomp})). To evaluate
$\rho_{\mbox{\tiny{GWB}}}$, introduce an Earth-centered Cartesian
coordinate system, described by unit basis vectors $\widehat{x}$,
$\widehat{y}$, and $\widehat{z}$, with 
\begin{itemize}
\item $\widehat{z}$ pointing parallel
to the Earth's polar axis and in the direction of the north celestial
pole, 
\item $\widehat{x}$ parallel to the line that runs in the equatorial
plane from Earth's center to the intersection of the equator with the
prime meridian at Greenwich, and 
\item $\widehat{y}$ chosen to form a
right-handed coordinate system.
\end{itemize}
Similarly, we introduce the usual spherical-polar coordinate system,
\begin{eqnarray}
r^2        & = &  x^2 + y^2 + z^2 \, , \\
\cos\theta & = &  z/r \, , \\
\tan\phi   & = &  y/x \, .
\end{eqnarray}
In these coordinates we write the gravitational wave polarization vectors 
as
\begin{eqnarray}\label{eq:epsilon}
\epsilon^+_{ab}
  & = &  \widehat{m}_a\widehat{m}_b-\widehat{n}_a\widehat{n}_b \, , \\
\epsilon^\times_{ab}
  & = &  \widehat{m}_a\widehat{n}_b+\widehat{n}_a\widehat{m}_b \, ,
\end{eqnarray}
where\footnote{Our conventions follow those of \citet{AlRo:99} except that 
we use $\widehat{\Omega}$ to denote the direction \emph{to} the GRB/GWB source 
on the sky, rather than the propagation direction of the GWB.  The net 
effect is $\widehat{m}\to-\widehat{m}$ in (\ref{eq:m}).}
\begin{eqnarray}
\widehat{m}
  & = &  -\widehat{x}\sin\phi+\widehat{y}\cos\theta \label{eq:m} \, , \\
\widehat{n}
  & = &  \widehat{x}\cos\phi\cos\theta +
\widehat{y}\sin\phi\cos\theta-\widehat{z}\sin\theta
         \, , \label{n} \\
\widehat{\Omega}
  & = &  \widehat{x}\cos\phi\sin\theta +
\widehat{y}\sin\phi\sin\theta + \widehat{z}\cos\theta
         \label{Omega} \, . 
\end{eqnarray}
Similarly, denoting the detector projection tensor (cf.\ equation 
(\ref{eq:dIFO})) for the LIGO Hanford (LIGO Livingston) Observatory
detector by $d_{\mbox{\tiny{LHO}}}^{ab}$ ($d_{\mbox{\tiny{LLO}}}^{ab}$)
we have \citep{AlHaJoLaWe:00}
\begin{eqnarray}
d_{\mbox{\tiny{LHO}}}^{ab}  & = &  \left(\begin{array}{rrr}
                     -0.3926 & -0.0776 & -0.2474 \\
                     -0.0776 &  0.3195 &  0.2280 \\ 
                     -0.2474 &  0.2280 &  0.0731 
                 \end{array}\right) \, , \\
d_{\mbox{\tiny{LLO}}}^{ab}  & = &  \left(\begin{array}{rrr}
                      0.4113 &  0.1402 &  0.2473 \\
                      0.1402 & -0.1090 & -0.1816 \\
                      0.2473 & -0.1816 & -0.3022
                 \end{array}\right) \, . 
\end{eqnarray}

\begin{figure}
  \begin{center}
  \includegraphics[height=8cm]{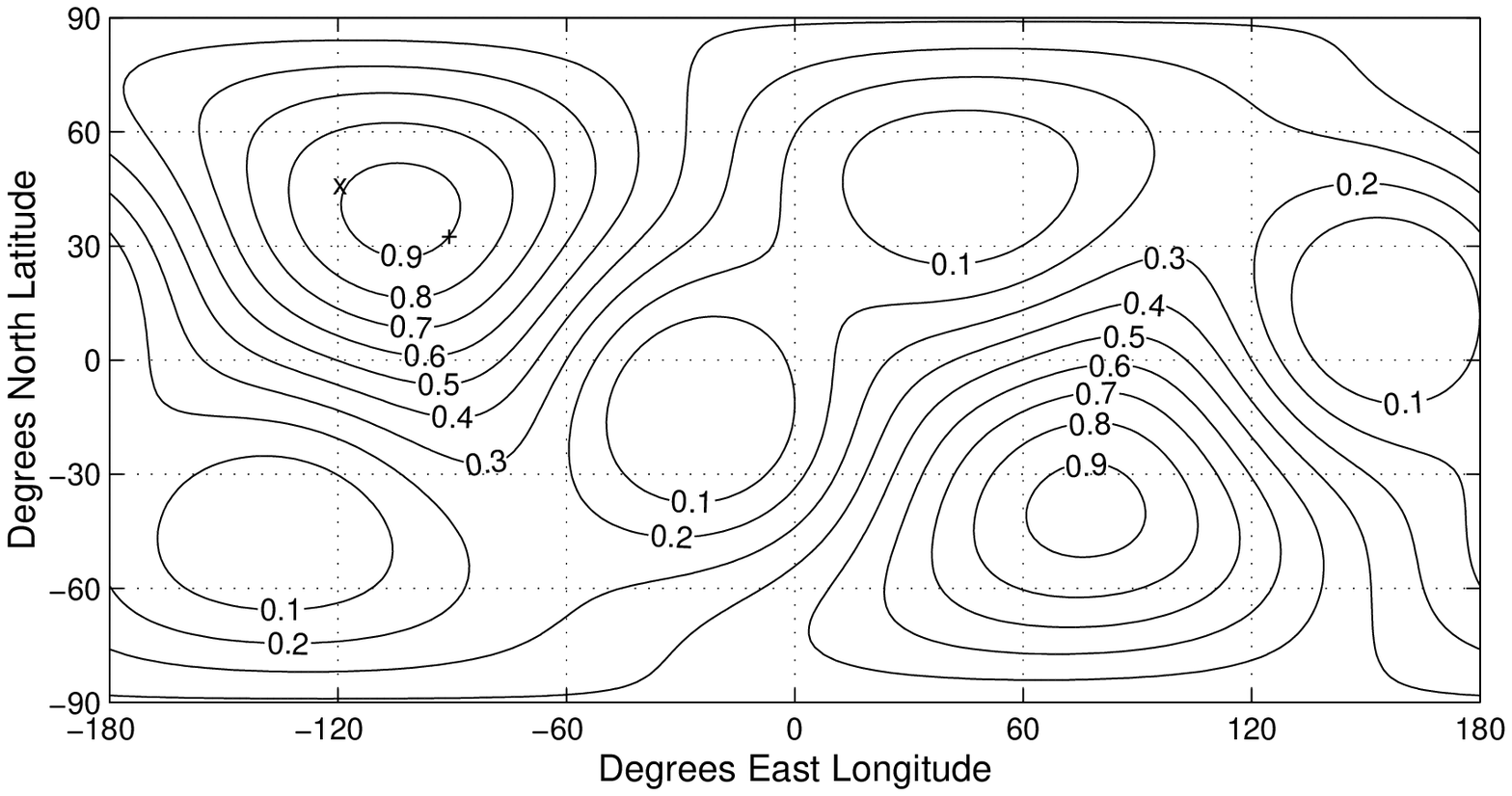}
  \caption{
\label{fig:LIGO}
LIGO sensitivity pattern $\rho_{\mbox{\tiny{}GWB}}$ 
(\ref{eq:rhoLIGO}) in Earth-based coordinates
for gravitational-wave bursts satisfying (\ref{eq:power}),
(\ref{eq:uncorrel}).  
The $+$ and $\times$ mark the locations of the LLO and LHO detectors.
The array is most sensitive in the directions
orthogonal to the plane of the LIGO detectors, corresponding to the two
peaks at upper left and lower right.  The sensitivity
is lowest in the plane of the detectors near the directions at 45
degrees to the arms and vanishes in certain directions, producing the
four wells of low sensitivity.  The sensitivity has been scaled to
range from $[0,1]$ in this plot.}
  \end{center}
\end{figure}

Figure~\ref{fig:LIGO} shows the antenna pattern
$\rho_{\mbox{\tiny{}GWB}}$ (\ref{eq:rhoLIGO}). Since the two LIGO
detectors share nearly the same plane and have arms nearly aligned
with each other their combined antenna pattern is very similar to that
for a single interferometer.  In particular, the LHO/LLO detector
combination is most sensitive to radiation arriving from a direction
orthogonal to the (nearly common) plane of the detector arms
(corresponding to the two peaks in Figure~\ref{fig:LIGO}) and
least sensitive to radiation arriving in the detector arm plane and
parallel to the (nearly common) arm bisector (producing the four wells
of low sensitivity in Figure~\ref{fig:LIGO}). It is also precisely
zero in certain directions.

\begin{figure} 
  \begin{center}
  \includegraphics[height=8cm]{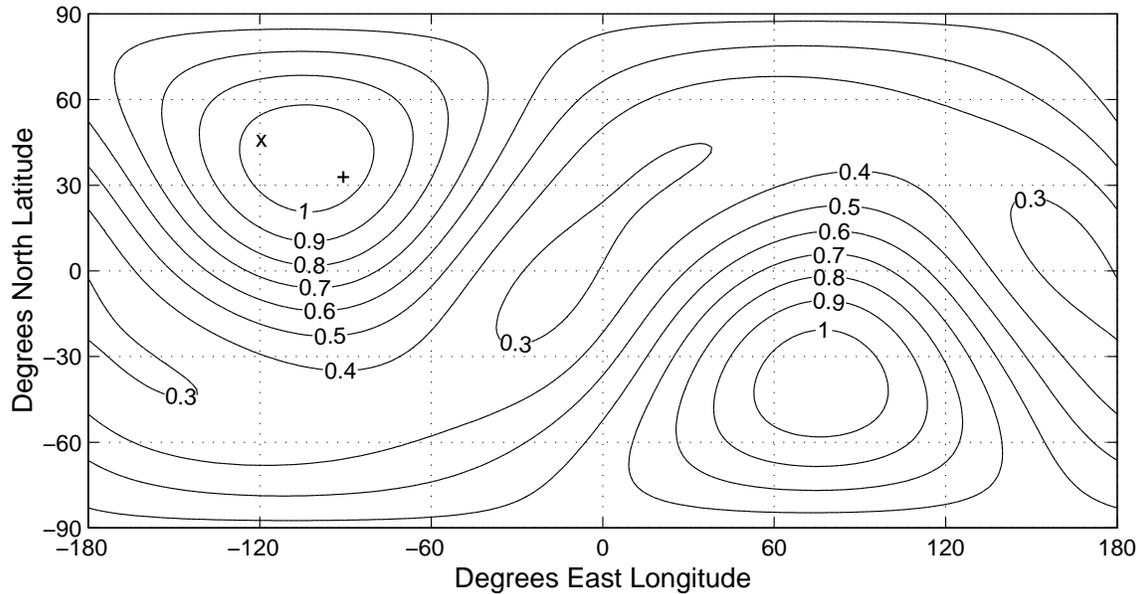}
  \caption{
\label{fig:FOM} 
Figure of merit $\zeta$ (\ref{eq:FOM}) for Swift pointing in Earth-based
coordinates, produced by convolving the LIGO sensitivity pattern
$\rho_{\mbox{\tiny{}GWB}}$ (\ref{eq:rhoLIGO}) with the Swift sensitivity
function $\rho_{\mbox{\tiny{}Swift}}$ (\ref{eq:rhoSwift}).  The four
minima of the LIGO sensitivity pattern of Figure~\ref{fig:LIGO} are
smeared into two minima which are not so deep.  The
figure of merit is nowhere zero, having a range of $[0.25,1.00]$ and an
all-sky average of 0.56.
The $+$ and $\times$ mark the locations of the LLO and LHO detectors.
}
  \end{center}
\end{figure}

Convolving $\rho_{\mbox{\tiny{}GWB}}$ with the Swift sensitivity
function $\rho_{\mbox{\tiny{}Swift}}$ as in equation (\ref{eq:FOM}) gives the
figure of merit $\zeta$ for the Swift pointing, which is shown in
Figure~\ref{fig:FOM}.  This plot is a smeared version of the LIGO
antenna pattern of Figure~\ref{fig:LIGO}.  In particular, the four
minima of the LIGO sensitivity are smeared into two minima which are
wider but not as deep.  The figure of merit is nowhere zero, varying
by a factor of approximately 4 between best (near zenith of detectors)
and worst (near planes of detectors at 45 degrees from arms)
orientations of Swift.  The all-sky average of the figure of merit is
0.56 times the maximum value.

\section{Discussion}

The currently accepted model for gamma-ray burst phenomena involves
the violent formation of a rapidly rotating approximately solar mass
black hole.  Gravitational waves should be associated with the black-hole 
formation, and their detection would permit this model to be
tested, the black-hole progenitor (e.g., coalescing binary or collapsing 
stellar core) to be identified, and the origin of the
gamma rays (within the expanding relativistic fireball or at the
point of impact on the interstellar medium) to be located. 
Even upper limits on the gravitational-wave strength associated 
with gamma-ray bursts could constrain the gamma-ray burst model.

We have evaluated how the quality of an upper limit on the
gravitational-wave strength associated with gamma-ray burst
observations depends on the relative orientation of the gamma-ray
burst and gravitational-wave detectors, with particular application to the
Swift Burst-Alert Telescope (BAT) and the LIGO gravitational wave
detectors. Setting aside other physical and science constraints on the
Swift mission, careful choice of BAT pointing leads to an upper limit
on the observed GRB population-averaged mean-square gravitational-wave
strength a factor of two lower than the upper limit resulting from
pointing that does not take this science into account.

There are, of course, numerous science and technical constraints that
determine the pointing profile of a satellite like Swift;
correspondingly, even in the most optimistic case the ratio of the
best possible upper limit to the best attainable upper limit will be
reduced from this factor of two. Nevertheless, when it can be done
without jeopardizing other mission objectives there is an advantage to
optimizing the pointing of Swift to maximize the joint LIGO/Swift
sensitivity to gamma-ray burst systems. The pointing dependent part
(cf.\ equation (\ref{eq:FOM})) of the anticipated upper limit that can be
set with these joint observations, suitably normalized, is closely
related to the science enabled by joint LIGO/Swift observations and
constitutes an excellent ``figure-of-merit'' that can be used to
incorporate this objective in Swift mission planning.

\acknowledgments

We are grateful to Margaret Chester, Shiho Kobayashi, 
and John Nousek for useful discussions.
LSF and PJS acknowledge NSF grant PHY~00-99559, and LSF, PJS and BK
acknowledge the Center for Gravitational Wave Physics, which is
supported by the NSF under cooperative agreement PHY~01-14375 and BK
acknowledges the support of the Albert Einstein Institut.





\end{document}